\documentclass[12pt]{iopart} 
\pdfoutput=1
\usepackage{graphicx} 
\usepackage[caption=false]{subfig}
\usepackage{color}
\usepackage{float}

\usepackage{lineno}
\usepackage{setspace}

\begin{document} 

\title[]{Growth-rate distributions of gut microbiota time series
}

\author{ E. Brigatti$^{1}$ and S. Azaele$^{2,3}$} 
\address{$^{1}$ 
Instituto de F\'{\i}sica, Universidade Federal do Rio de Janeiro,  Av. Athos da Silveira Ramos, 149, Cidade Universit\'aria, 21941-972, Rio de Janeiro, RJ, Brazil}
\address{$^{2}$ 
Dipartimento di Fisica  ``G. Galilei", Universit\`a di Padova , via Marzolo 8, 35131 Padova, Italy.}
\address{$^{3}$ 
INFN, Istituto Nazionale di Fisica Nucleare, 35131, Padova, Italy.}

\ead{edgardo@if.ufrj.br}


\maketitle

\begin{abstract} 

Logarithmic growth-rates are fundamental observables for describing ecological systems and the characterization of their distributions with analytical techniques can greatly improve their comprehension. 
Here a neutral model based on a stochastic differential equation with demographic noise, which presents a closed form for these distributions, is used to describe the population dynamics of microbiota.
Results show that this model can successfully reproduce the log-growth rate distribution
of the considered abundance time-series. 
More significantly, it predicts its temporal dependence, by reproducing its kurtosis evolution when the time lag $\tau$ is increased. Furthermore, its typical shape for large $\tau$ is assessed, verifying that the distribution variance  does not diverge with $\tau$.
The simulated processes generated by the calibrated stochastic equation and the analysis of each time-series, taken one by one, provided additional support for our approach.

Alternatively, we tried to describe our dataset by using a logistic neutral model with an environmental stochastic term. Analytical and numerical results show that this model is not suited for describing the leptokurtic log-growth rates distribution found in our data.
These results support an effective neutral model with demographic stochasticity for describing 
the considered microbiota. 

\end{abstract} 



\section{Introduction}



Quantitative studies of gut microbiota have generated an increasing and rich 
body of information 
about different features of these ecosystems.
This has been made possible by 
the fact that nowadays these 
microbial communities present a more 
experimental accessibility than other macroscopic
ecological communities, as current sequencing technologies
can monitor with high temporal resolution
the dynamics of abundance of bacteria \cite{Faust}.
Numerous studies have described some aspects of 
the community ecology of these systems.
They outlined macroecological patterns that characterize global statistical relationships
of spatial and temporal variation and taxonomical
diversity \cite{Li16,Shoemaker,Grilli0} 
and even their  temporal changes associated with host health, diet and lifestyle \cite{David}.
However, a robust and general description of gut microbiota at the level of population
dynamics is still missing.  
Therefore, it is interesting and important to identify quantitative statistical stylized 
facts 
which can characterize the 
population dynamics of gut bacteria, with the ultimate goal of 
pointing out the biological processes and factors governing these dynamics. 

In this work we will focus on the population dynamics displayed by the 
different species of bacteria present in gut  microbiota.
A very important observable used 
in the description of population dynamics is the logarithmic growth rate.
Given an abundance time series $x(t)$, the log-growth rate is
defined as: $g(t,\tau)=\ln{x(t+\tau)}-\ln{x(t)}$.
The time series of this quantity have been traditionally used in population dynamics
for analyzing important features of the corresponding populations. 
Some analysis methods proposed by Royama \cite{Royama} 
are capable of generating clear diagnoses of 
their ecological features, 
in particular density dependence.
This phenomenological approach 
is based on the 
fit of the population growth rate and it is more efficient in testing  
and diagnosing, rather than in modeling \cite{Berryman,Brigatti}. 

The idea behind the use of this quantity, instead of directly considering
abundance time series, is linked to  the 
construction of an observable close to stationarity.
In fact, abundances are frequently non-stationary and subject to 
large variations  and measurement errors.
It follows that it is more robust 
to consider the fluctuations of the observable, measured as 
the difference between two consecutive logarithms of the abundance.
Indeed,  if the process has
multiplicative dynamics, logarithms are a natural choice because 
they eliminate underlying exponential growth trends. 
Finally, it is interesting to note that this observable has been 
traditionally associated with a 
discrete approximation for the instantaneous population per capita growth rate, since $\ln(x(t + 1))-\ln (x(t))\approx \frac{1}{x}\frac{dx}{dt}$ in a first, rough approximation. 

In this study we will focus on the characterization of the $P(g,\tau)$, the  
distribution of the $g(t,\tau)$ when the process has reached stationarity,
and therefore $g$ does not depend on $t$.
This distribution has been already studied for the microbiota by Ji {\it et al.} \cite{Ji}.
Moreover, it has been
actively investigated not only in other ecological systems \cite{Keitt98}, but also in finance (the return distribution) \cite{Cont}, 
economics \cite{Stanley,Dosi,Schwarzkopf} and social science \cite{Luiz}. 
There, the considered underlying observables were not population abundances
but, respectively, prices, firms or fund sizes, and crimes.

Early simple multiplicative models, discrete or continuous,
hypothesized that the Gaussian distribution should emerge as a typical shape
modeling this distribution. 
In contrast,  empirical studies showed
that, generally, $P(g,\tau)$ are  not simply characterized by Gaussian distributions,
but it is common to find shapes close to the Laplace distribution \cite{Keitt98,Stanley,Luiz}
or, in finance, distributions with power-law tails \cite{Cont,DeSouza}.
There is a second, important, statistical feature characterizing
the temporal dependence of the shape of $P(g,\tau)$. 
$P(g,\tau)$ is markedly dependent on $\tau$ and seems to be
attracted towards a Gaussian shape for increasing values of this parameter.
This last behavior is well known in finance \cite{Cont,Schwarzkopf},  
where it is called aggregational gaussianity
and it is justified on the basis of the central limit theorem. 
As $g(t,\tau)$ is time additive, for independent  data points, 
the distribution is expected to converge to a Gaussian one for large $\tau$.\\

In this study we will characterize $P(g,\tau)$ and its temporal dependence by
using a stochastic model which describes the underlying abundance 
time series \cite{Azaele}.  This method allows to analytically obtain 
the $P(g,\tau)$ and its temporal dependence
and to compare them with the empirical data.
In addition to the description of the principal features of the 
log-growth rate distribution, this approach suggests 
a specific model of population dynamics for the species present in the considered 
microbiota.
This model is part of a class of models which describe ecological systems 
within a neutral framework.
Neutral theories \cite{Hubbel,Volkov} posit that the dominant factors that determine the structures of an ecological community are driven by the demographic randomness
present in the populations,
which determine their random drift. By contrast,
the selection produced by the interactions among the individuals, the species identity and the environment effects are considered far less relevant.
In its more universal  
implementation, the neutral theory of biodiversity models the organisms of a community with identical per capita death, birth, immigration and speciation rates \cite{Hubbel}. 
Species are considered demographically and ecologically equivalent
and characterized by the same demographic  rates. 
Among the different models generated by these ideas, we consider a very simple and general one, based on a stochastic differential equation (SDE) 
which describes the dynamics of the population.
It is driven by a  linear drift and the noise term includes the square root of the population, 
which describes demographic noise \cite{Azaele}.
This approach generates predictions at stationarity,
which have been successfully applied in a variety of different systems, such as \cite{Sandro}.   
Indeed, this SDE is a paradigmatic neutral model which can be solved analytically and can be benchmarked under different conditions. 
At stationarity, it predicts a distribution of species abundances which is in good agreement with empirical data collected in neotropical forests 
and coral reefs \cite{Volkov,Sandro}.
The temporal dynamics of species abundances is also well captured \cite{Azaele}
and can be used to predict characteristic temporal scales which are empirically inaccessible. Finally, the model is amenable to spatial generalisations \cite{Peruzzo}
which generate patterns across scales very close to those observed empirically.
More recently, it has been showed that it is able to capture universal features of fluctuations in disparate systems \cite{Ashish}, thus improving our ability to forecast rare events.

With the aim of introducing some comparisons between the statistical patterns generated 
by different approaches, we consider a second 
model recently used for describing population dynamics in microbiota \cite{eLife,Grilli}. 
In this approach, populations are modeled by a traditional logistic growth term, coupled 
with a source of environmental stochasticity implemented by a simple multiplicative term.
In section 3 these two models will be described in detail.




\section{Data}



In recent times microbiome data have become increasingly accessible and a variety of different datasets have been analyzed in several studies.
Here we focus on the dataset previously considered by  Zaoli {\it et al.} \cite{Grilli}.
As we are interested in the population dynamics 
of the bacterial abundance, and not in the characterization of the community ecology, 
we look for data focusing on their statistical quality 
(relatively high sampling frequencies and relatively few gaps), rather than worrying about their generality.
For this reason,  
we select time-series presenting daily sampling frequency, 
and which display positive read counts
in at least $75\%$ of the data points.
We end up with data from four healthy human individuals: 
2 individuals, M3 and F4, from the Moving Pictures MP dataset \cite{Caporaso}
(indicated as M3F4 reads), 
and the time-series of the post-travel period of individual A and the pre-Salmonella interval of individual B of the study of David {\it et al.} \cite{David} (indicated as IndAB reads).
This dataset corresponds to a total of 1305 abundance time-series of 
bacterial operational taxonomical units (OTU)
and present a length spanning from just over 4 months to around 14.
There is a huge variability in their abundances. 
Examples of some time-series can be seen in Fig.\ref{Fig_tauDep}.
These time-series are not all stationary.
The statistical properties of stationary time-series are independent of the point in time at which they are observed. In fact, in these series, the joint probability distribution of any subsequence of data points 
does not change with a shift in time.
As we will focus our analysis on time-independent distributions and the analytical distributions obtained from the considered theoretical models are valid only at stationarity,
we restrict our dataset to the stationary time-series.
These series are selected by using the Dickey-Fuller test \cite{DF} and
correspond to 75\% of the original dataset.




\section{Methods}


The neutral framework can be implemented by describing the abundance
dynamics $x(t)$ with the following SDE:
\begin{eqnarray}
dx_t= (b-x_t/a)dt+\sqrt{2D x_t}dW_t,
\label{SDE}
\end{eqnarray}
where $a,b,D>0$ and $W_t$ is a standard Wiener process.  
The solution of this equation is a randomly fluctuating
population which is 
drawn back to a long-term deterministic value equal to $ab$, with 
a correlation time determined by $a$ and fluctuations controlled by the parameter $D$. 
The process tends to drift towards its long-term stationary value because of its deterministic decaying term. The return time at which populations come back to the stationary abundance after a disturbance is equal to $a$.
This SDE is known in the literature as the Cox-Ingersoll-Ross (CIR) equation.
It was first introduced by Feller for modeling population growth \cite{Feller} and, 
following its use to model interest rates \cite{CIR},
it became very popular in the finance literature. 
More recently, it has been adopted for describing Neutral dynamics in ecological systems \cite{Azaele} and we consider this interpretation as the theoretical underpinning 
of its use in this work.
This equation has an explicit solution.
The propagator 
can be also analytically obtained and
the stationary distribution 
is given by the Gamma distribution:

\begin{equation}
P(x)=
\frac{x^{\alpha-1}e^{-x/Da}}{\Gamma(\alpha)(Da)^{\alpha}},
\label{ana_abundance}
\end{equation}

where we introduce the notation $\alpha=b/D$.
This closed form allows to directly compare the empirical abundance distribution obtained from stationary time-series and the one generated by the considered SDE.
Despite this process has been frequently used in finance, where returns are a fundamental measure, only recently an analytical result have been obtained for  describing
the log-growth rate distributions $P(g,\tau)$ \cite{Azaele}.
Following Azaele {\it et al.} \cite{Azaele}, this distribution can be obtained
starting from 
the conditional transition density
and the stationary probability distribution, 
which allows to calculate the probability for a given time lag $\tau$ of the ratio 
$r(\tau)=x(t_0+\tau)/x(t_0)$, where $x(t_0+\tau)$, with $\tau>0$, and $x(t_0)$ are 
the population abundance at time $t_0+\tau$ and $t_0$, respectively.
Assuming stationarity at $t_0$, the probability distribution
of $r(\tau)$ does not depend on $t_0$.
The logarithmic growth rate distribution is finally obtained 
after the change of variable $g(\tau)=\ln{r(\tau)}$ and is given by:


\begin{equation}
P(g,\tau)=C
\frac{e^{g}+1}{e^{g(\alpha+1)}}
\bigg[\frac{4e^{2g}}{(e^{g}+1)^2e^{\tau/a}-4e^{g}}\bigg]^{\alpha+1/2}
\bigg[{\sinh(\tau/2a)}\bigg]^{\alpha+1}
\frac{e^{\frac{\alpha\tau}{2a}}}{1-e^{-\tau/a}}
\label{ana_growth}
\end{equation}
where $C$ is a renormalization constant equal to 
$\frac{2^{\alpha-1}}{\sqrt{\pi}}\frac{\Gamma(\alpha+1/2)}{\Gamma(\alpha)}$.
Note that the asymptotic behavior of 
the tails of the distribution, for $g \to \pm \infty$ 
and fixed $\tau$, follow $e^{-\alpha g}$. 

 For $\tau \to \infty$ eq. \ref{ana_growth} reduces to:
\begin{equation}
P(g)=C 
\frac{2^\alpha e^{\alpha g}}{(e^{g}+1)^{2\alpha}}
\label{ana_growthLong}
\end{equation}
which, remarkably, is not dependent on $a$. 
Note that for $\alpha=1$, it is a logistic distribution, 
a well known distribution with a shape quite similar to the normal one, 
but with 
heavier tails.\\

An alternative model which has been recently used for describing
the abundance dynamics of microbiota \cite{eLife,Grilli,Grilli2} 
is the logistic model, which is described by the following SDE:
\begin{eqnarray}
dx_t= \frac{x_t}{a}(1-x_t/K)dt+\sqrt{\frac{\sigma^2}{a}} x_t dW_t,
\label{eq_Log}
\end{eqnarray}

If $\sigma^2<2$, the stationary distribution of this SDE also follows a Gamma law \cite{Grilli}, with mean $\langle x \rangle =  K(1-\sigma^2/2)$.
Unluckily, the propagator \cite{Otunuga} cannot be used to calculate the log-growth distribution.
In this case we can grasp some information about the behavior of
 $P(g,\tau)$ 
by calculating $d \ln{\frac{x}{\langle x \rangle}}$.
By defining  $y=\ln{\bigg(\frac{x}{K(1-\sigma^2/2)}\bigg)}$ and   following   the It\^{o} calculus, 
we obtain $dy=\frac{1}{x}dx-\frac{1}{2x^2}(dx)^2$, from which it follows:

\begin{eqnarray}
dy_t=\frac{1}{a}(1-\sigma^2/2)(1-e^{y_t})dt+\sqrt{\frac{\sigma^2}{a}} dW_t. 
\label{eq_LogRet0}
\end{eqnarray}
This expression, for $y\sim0$, can be  linearly approximated by:
\begin{eqnarray}
dy_t\sim-\frac{1}{a}(1-\sigma^2/2)y_tdt+\sqrt{\frac{\sigma^2}{a}} dW_t,
\label{eq_LogRet1}
\end{eqnarray}
which is a Ornstein-Uhlenbeck process.
It presents the stationary distribution $\widetilde{P}_s(y)=N\bigg(0,\frac{\sigma^2}{2(1-\sigma^2/2)}\bigg)$ and also a Gaussian propagator $\widetilde{P}(y,t_0+\tau | y',t_0)$ with mean $y'e^{-\frac{\tau(1-\sigma^2/2)}{a}}$  and variance $\frac{\sigma^2(1-e^{-\frac{2\tau(1-\sigma^2/2)}{a}})}{2(1-\sigma^2/2)}$.

\begin{figure}[h]
\begin{center}
\includegraphics[angle=0,width=0.95\textwidth]{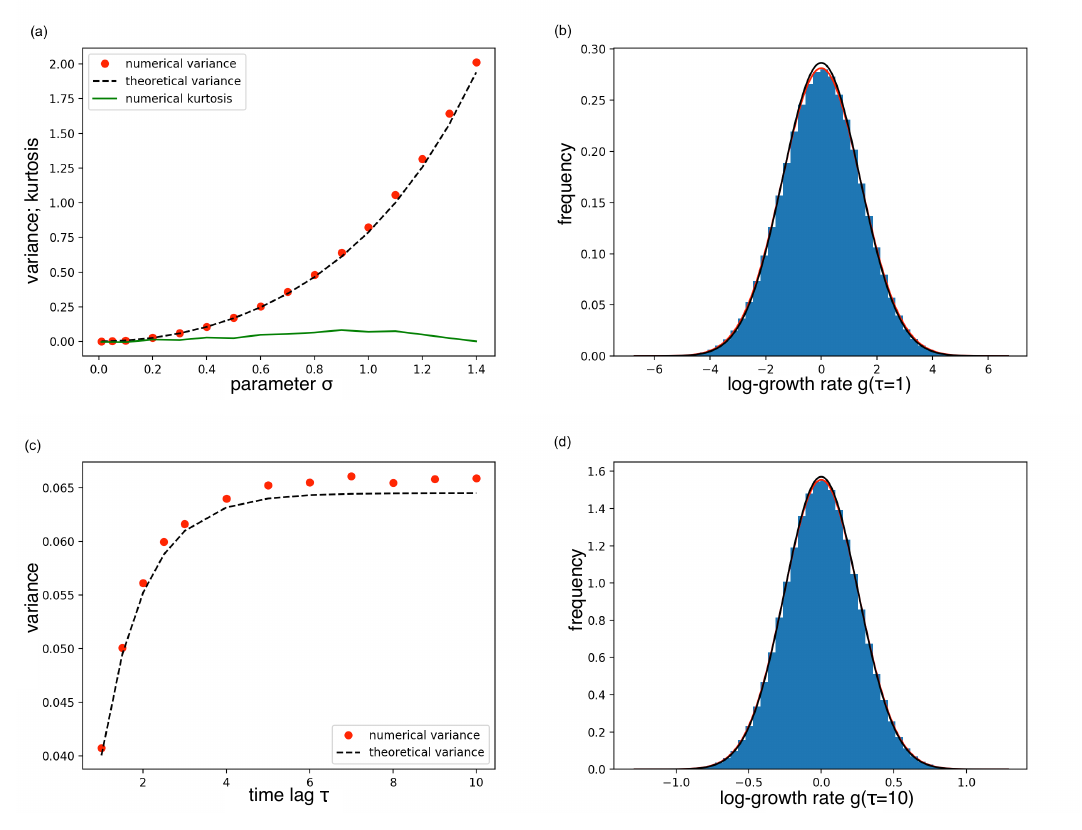}
\end{center}
\caption{\small 
(a) Variance of $P(g,\tau=1)$ as a function of $\sigma$. Red points are obtained
from numerical simulations of the SDE of eq. \ref{eq_Log} ($K=1,a=1$).
The black dashed line is the analytical approximation of eq.  \ref{eq_DisFin}.
The solid green line represents 
the excess kurtosis of the distributions obtained by the numerical simulations. 
They are always very close to zero, showing that the distributions have always 
 a Gaussian shape. 
(b) $P(g,\tau=1)$ for the numerical simulations with $K=1,a=1,\sigma=1.4$.
 The black line is the distribution obtained from the analytical
approximation, the red one is a Gaussian fitting.
Note that with these parameter values we are far from the regime where $P(y)$
of eq. \ref{eq_LogRet0} is gaussian, 
nevertheless the approximation continues to work very well.
(c) Variance of $P(g,\tau)$ as a function of $\tau$. Red points are obtained
from numerical simulations  ($K=1,\sigma=0.25$).  
The black dashed line is the analytical approximation.
(d) $P(g,\tau=10)$ for the numerical simulations ($K=1,a=1,\sigma=0.25$).  
The black line is the analytical approximation, the red one is a Gaussian fitting.}.
\label{Fig_theo}
\end{figure}

If $x(t)$ is the process of eq. \ref{eq_Log}, at stationarity
$P(g,\tau)=\langle  \delta(g-\ln{(\frac{x(t_0+\tau)}{x(t_0)}})) \rangle=\langle  \delta(g-(y-y')) \rangle$,
from which $P(g,\tau)=\int_{-\infty}^{+\infty} dy \int_{-\infty}^{+\infty} dy' \delta(g-(y-y')) P(y,t_0+\tau | y',t_0) P_s(y')$, where there is no dependence on $t_0$ because we assumed stationarity.
The approximation $y\sim0$ is satisfied if $\frac{x(t)-\langle x \rangle}{\langle x \rangle}\sim0$,
which is typically true if the coefficient of variation $\sqrt{var(x)/\langle x \rangle^2}$
is small. As the coefficient of variation of the process in eq.\ref{eq_Log} depends only on $\sigma$, this approximation
is valid in the regime of small $\sigma$, regardless of the other parameters. 
Thus, in this regime, we can use the propagator and the stationary distribution from the SDE
of eq. \ref{eq_LogRet1} and calculate analytically the integral, obtaining:

\begin{eqnarray}
P(g,\tau) \approx N\bigg(0,\frac{\sigma^2(1-e^{-\frac{\tau(1-\sigma^2/2)}{a}})}{(1-\sigma^2/2)}\bigg),
\label{eq_DisFin}
\end{eqnarray}

where $N$ specifies 
a Normal distribution with the corresponding
mean and standard deviation.
Numerical simulations confirm this conclusion and
show that the approximation is fine even 
for all the allowed $\sigma$ values (see Fig. \ref{Fig_theo}).
For example, for $\sigma=1.4$, $P(g,\tau)$ is perfectly normal, as can be seen
by fitting it with a generic Gaussian 
distribution. The difference in the variance of this generic Gaussian and the one of equation \ref{eq_DisFin} is only $3.6\%$.
Finally, 
numerical simulations verify
that the shape of the $P(g,\tau)$, regardless of the regimes considered,
is always 
indistinguishable from a Gaussian one.


\section{Results}

\begin{figure}[h]
\begin{center}
\includegraphics[angle=0,width=0.46\textwidth]{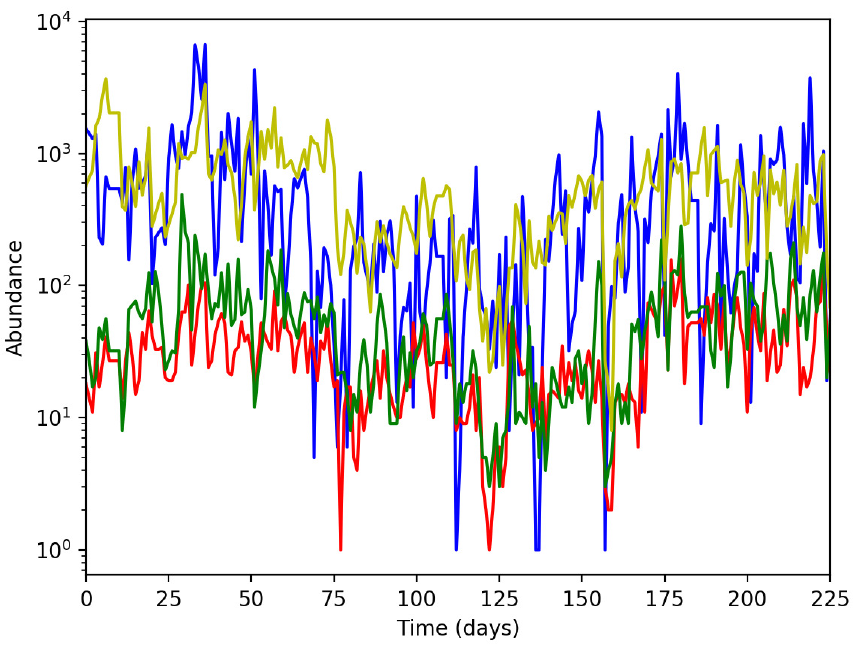}
\includegraphics[angle=0,width=0.43\textwidth]{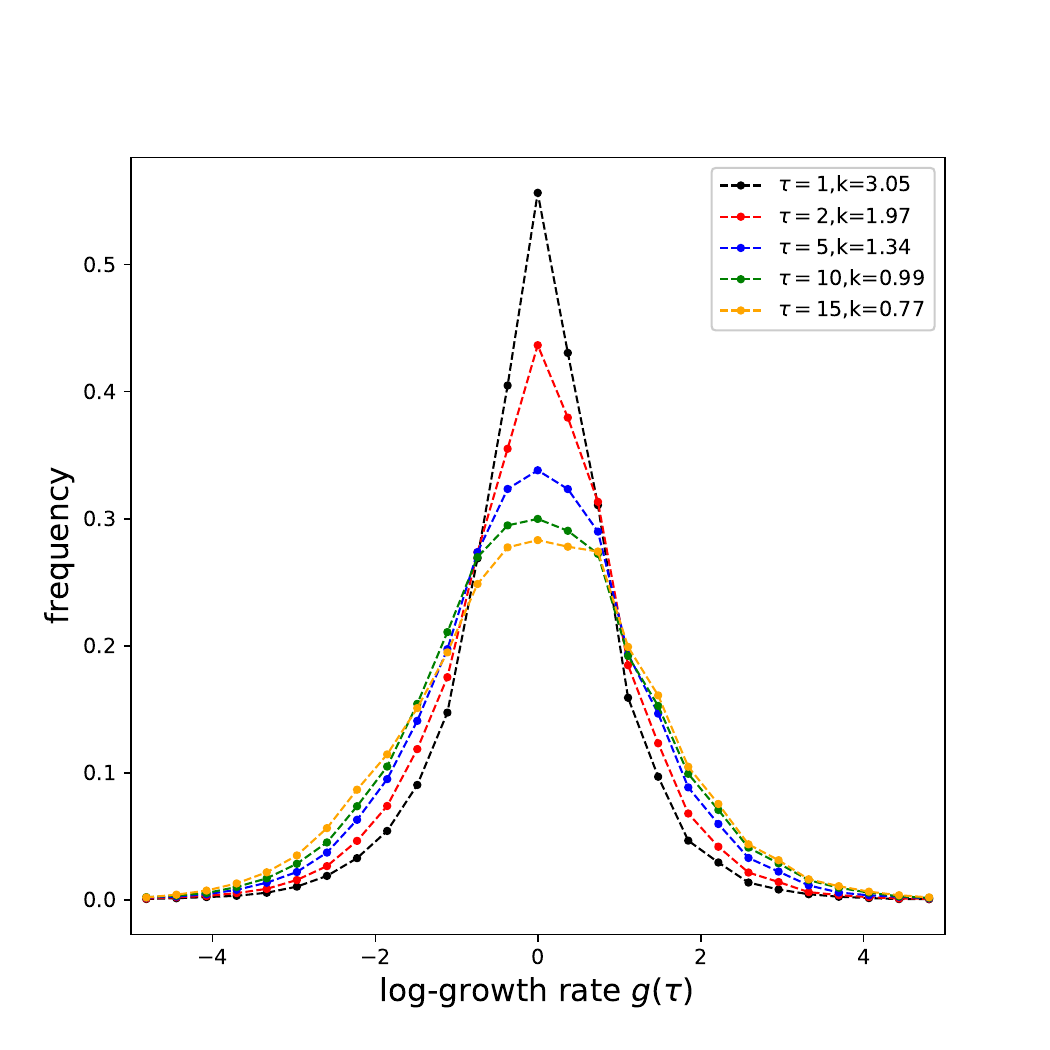}
\end{center}
\caption{\small 
On the left: examples of some empirical abundance time-series.  
We can note the wide fluctuations. Actually, the coefficient of variation is practically always close to or greater than one for all the considered time series.
On the right:
Distribution of empirical log-growth rates measured at different time lags $\tau$.
By increasing $\tau$, the distributions present lower peaks
but fewer outliers.
This change results in smaller excess kurtosis $k$.
}
\label{Fig_tauDep}
\end{figure}

Here we present the study of a selection of the dataset described
in section 2, analysing the IndAB time-series. 
However, similar results are obtained when considering the whole dataset (see
Supplementary Material).
The first part of the investigation analyzes the ensemble of all time-series 
present in this selected dataset together, assuming that the same single
 parametrization of the CIR model is able to describe the population dynamics 
 of all OTUs. 
 Supposing that all OTUs do not strongly interact between each other,
 we can model them as a community with effective parameters, in which OTUs behave similarly.
Although this assumption is only a rough approximation, the considered 
time series of the single OTUs can still be described by a distribution that captures the behavior of the whole community, which is in a regime of large fluctuations. This modeling approach is also supported by the results that we will present in fig.6. 
For this reason, we can hypothesize 
 that the mixture  of the single OTUs is well described 
 by the neutral distributions  for the whole community with effective parameters. 
Thus, for the analysis of the abundance distribution we merge the abundances of all the different OTUs present in the IndAB time-series.
In the case of the log-growth rate distribution, we calculate the log-growth rates for each OTU present in the IndAB dataset and then merge the arrays produced for all OTUs into a single one, obtaining their distribution.

First, we look at the log-growth rate distributions $P(g,\tau)$ 
for different values of $\tau$ (see Fig. \ref{Fig_tauDep}).
For small  $\tau$ the distributions present a characteristic shape
with a sharp and high peak followed by
tails that decay slower than a Gaussian distribution.
This shape is well characterized by the excess kurtosis  ($k=\langle(\frac{x-\mu}{\sigma})^4\rangle-3$, where $\mu=\langle x \rangle$ and $\sigma^2=\langle(x-\mu)^2\rangle$), which is obviously positive and, for $\tau=1$, close to 3.
Increasing $\tau$ the excess kurtosis becomes noticeably smaller:
peaks decrease and tails become 
more similar to those of a Gaussian distribution.

\begin{figure}[h]
\begin{center}
\includegraphics[angle=0,width=0.44\textwidth]{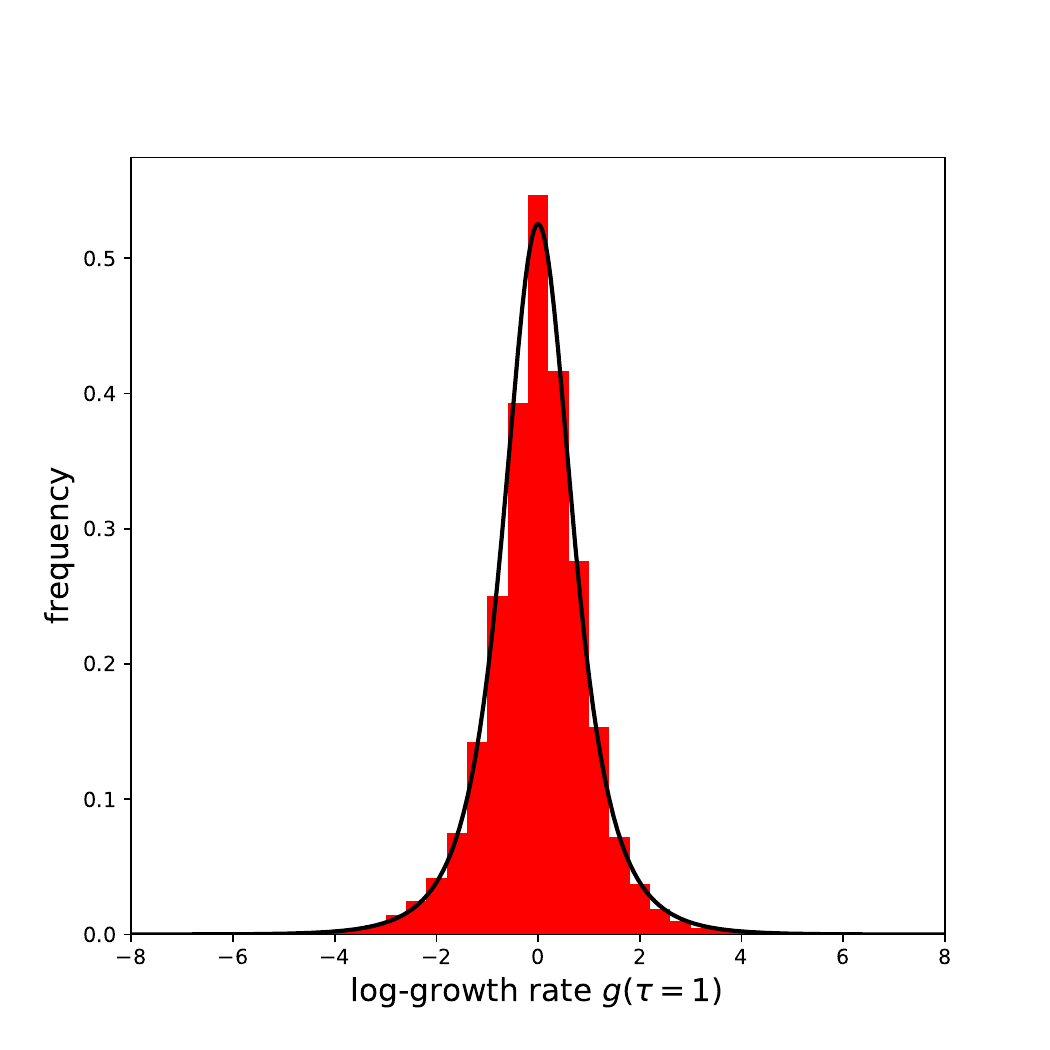}
\includegraphics[angle=0,width=0.45\textwidth]{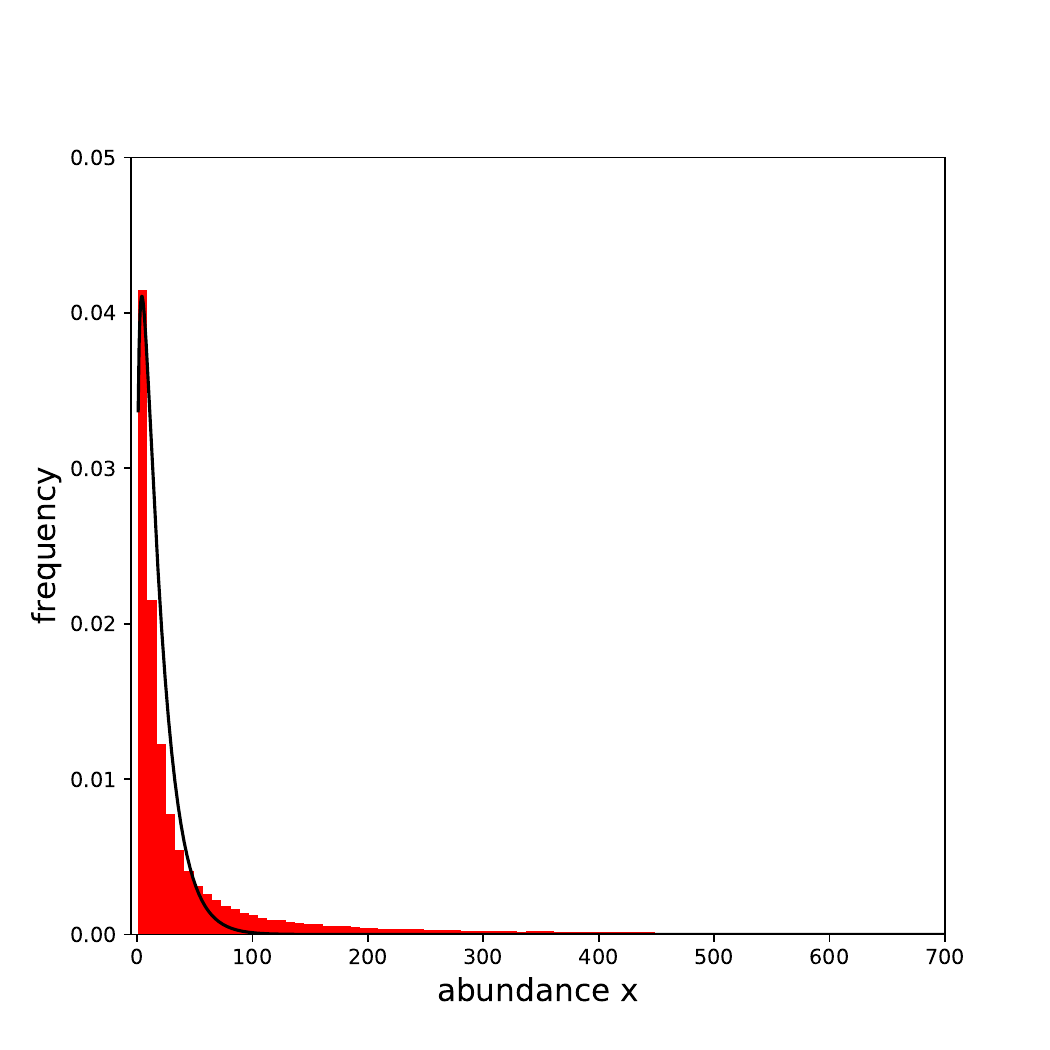}
\end{center}
\caption{\small 
On the left: fit of $P(g,\tau=1)$ using the analytical expression of eq. \ref{ana_growth}. 
On the right: the stationary abundance distribution fitted using eq. \ref{ana_abundance}
once fixed $\alpha$
with the results obtained by the estimation of the $P(g,\tau=1)$ parameters.
The solid black lines represent the fits, the red histograms show the empirical data.
} 
\label{Fig_growthFit}
\end{figure}

We focus on $P(g,\tau=1)$ and fit this distribution with eq. \ref{ana_growth}.
Parameters are estimated using the Maximum Log-likelihood method, 
which provides the values $\alpha=b/D=1.30\pm0.01$ and $a=2.70\pm0.02$.
The fit is excellent, as can be visually confirmed in Fig. \ref{Fig_growthFit}.
We  numerically computed the excess kurtosis of the fitted distribution, obtaining a value of 3.00, which very well approximates the empirical one (3.05).
Note that the model described by equation \ref{eq_Log} can not reproduce the empirical $P(g,\tau=1)$. This is because 
it can generate only distributions with negligible excess  kurtosis (Gaussian distributions), while the empirical one is positive and large.  
Since in our study we are interested in identifying a departure of $P(g,\tau)$ from a Gaussian shape 
focusing on the behavior of the tails, the analysis of the kurtosis
is a sufficient test for identifying non-normality.
We have also performed statistical normality tests. The  Jarque-Bera and the Anderson-Darling tests give a practically null p-value, 
excluding the null hypothesis of normal distribution. 
A fit with a normal distribution provides a poor result as well
(see Supplementary material).

To find further support for our hypothesis, we fit  experimental abundance data
coming from all our time-series 
with the stationary distribution of eq. \ref{ana_abundance} once fixed $\alpha$
with the results obtained by the estimation of the $P(g,\tau=1)$ parameters.
We obtained $D\cdot a = 14.06\pm0.08$.
Gamma distributions have been already used for describing the abundance of microbiota  \cite{Grilli0,eLife,Grilli}. 
Bearing in mind that our fit was obtained after fixing $\alpha$, and therefore by varying only the parameter $D\cdot a$, we can consider its quality acceptable.

\begin{figure}[h]
\begin{center}
\includegraphics[angle=0,width=0.5\textwidth]{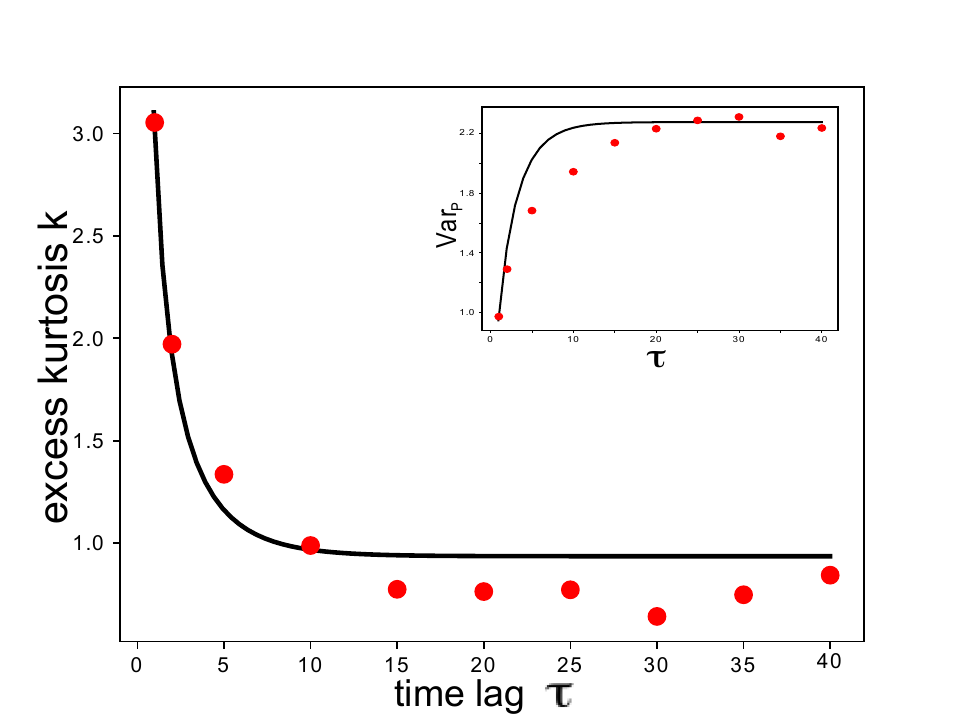}
\includegraphics[angle=0,width=0.4\textwidth]{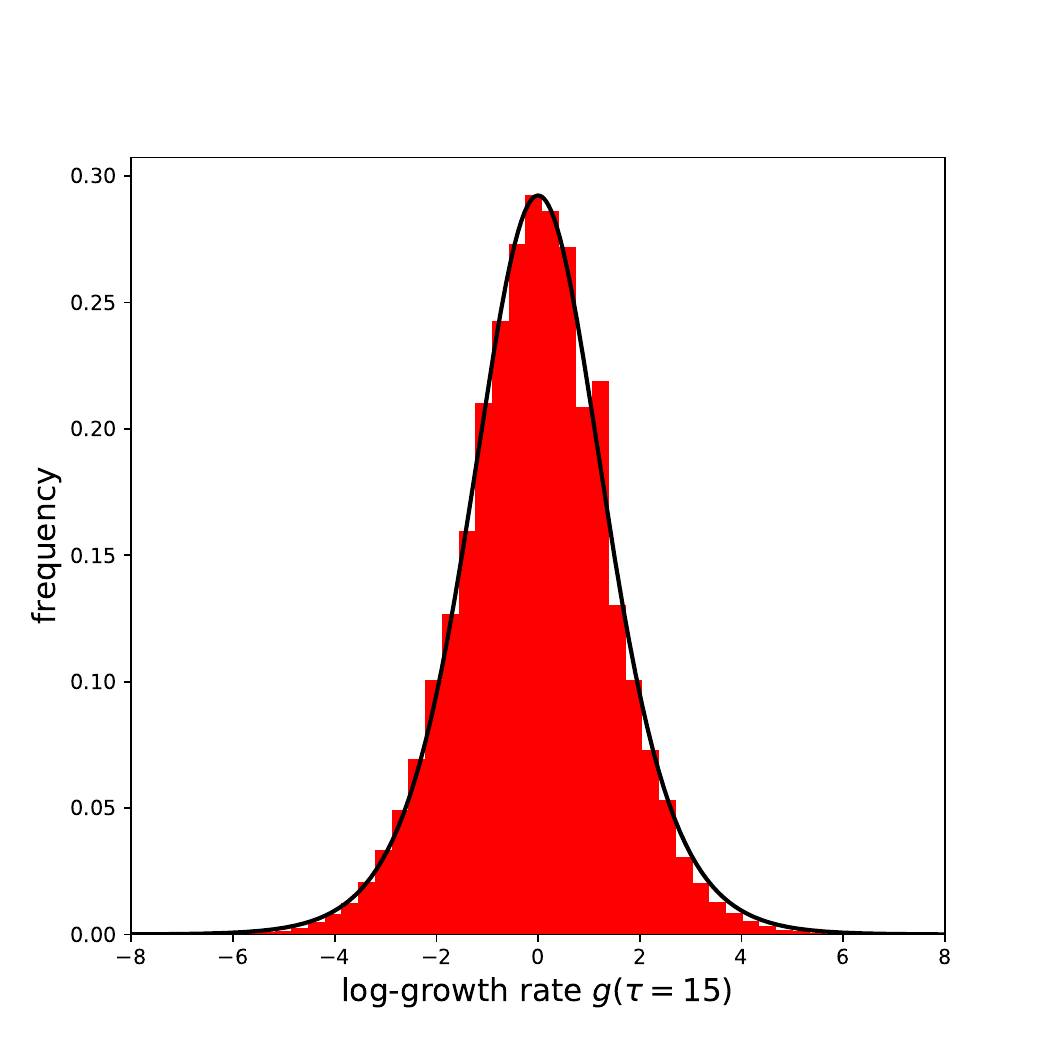}
\end{center}
\caption{\small 
On the left: excess kurtosis as a function of $\tau$. 
In the inset, lag-dependence of the variance of $P(g,\tau)$. 
The solid lines are the values predicted by using eq. \ref{ana_growth}, using 
the parameter values estimated at $\tau=1$. Red points are obtained from empirical data.
On the right: Solid line represents the analytical distribution for $\tau >>1$ (see eq. \ref{ana_growthLong}) when using $\alpha$ estimated at $\tau=1$. This curve  approximates well $P(g,\tau=15)$ obtained from the experimental dataset.}
\label{Fig_kurtosis}
\end{figure}

Next, we test how our model can well describe the temporal shift 
of the shape of the $P(g,\tau)$ 
towards a distribution with smaller excess kurtosis. 
We do that by comparing the analytical
prediction produced by eq. \ref{ana_growth} with our experimental data.
Predictions are generated calculating the excess kurtosis at different 
$\tau$ values from the analytical distribution with the parameters fixed by the estimation at 
$\tau=1$. Taking into account that kurtosis estimation is very sensitive to  outliers, the excess kurtosis obtained from empirical data are very well matched by the analytical prediction, as can be seen in Fig. \ref{Fig_kurtosis}.
In the same Figure,  
the variance of $P(g,\tau)$, indicated as Var$_P$, is displayed as a function of $\tau$. Empirical data grow  
and then saturate to a limit value. 
Predictions are produced using the same approach used for the excess kurtosis,
and capture very well the value reached at saturation.

By using the $\alpha$ value estimated at $\tau=1$  equation \ref{ana_growthLong} can predict the $P(g,\tau)$ for $\tau \to \infty$, which is a very good approximation for every distribution with $\tau >>1$.
The 
result is displayed in  Fig. \ref{Fig_kurtosis}.\\

\begin{figure}[h]
\begin{center}
\includegraphics[angle=0,width=0.4\textwidth]{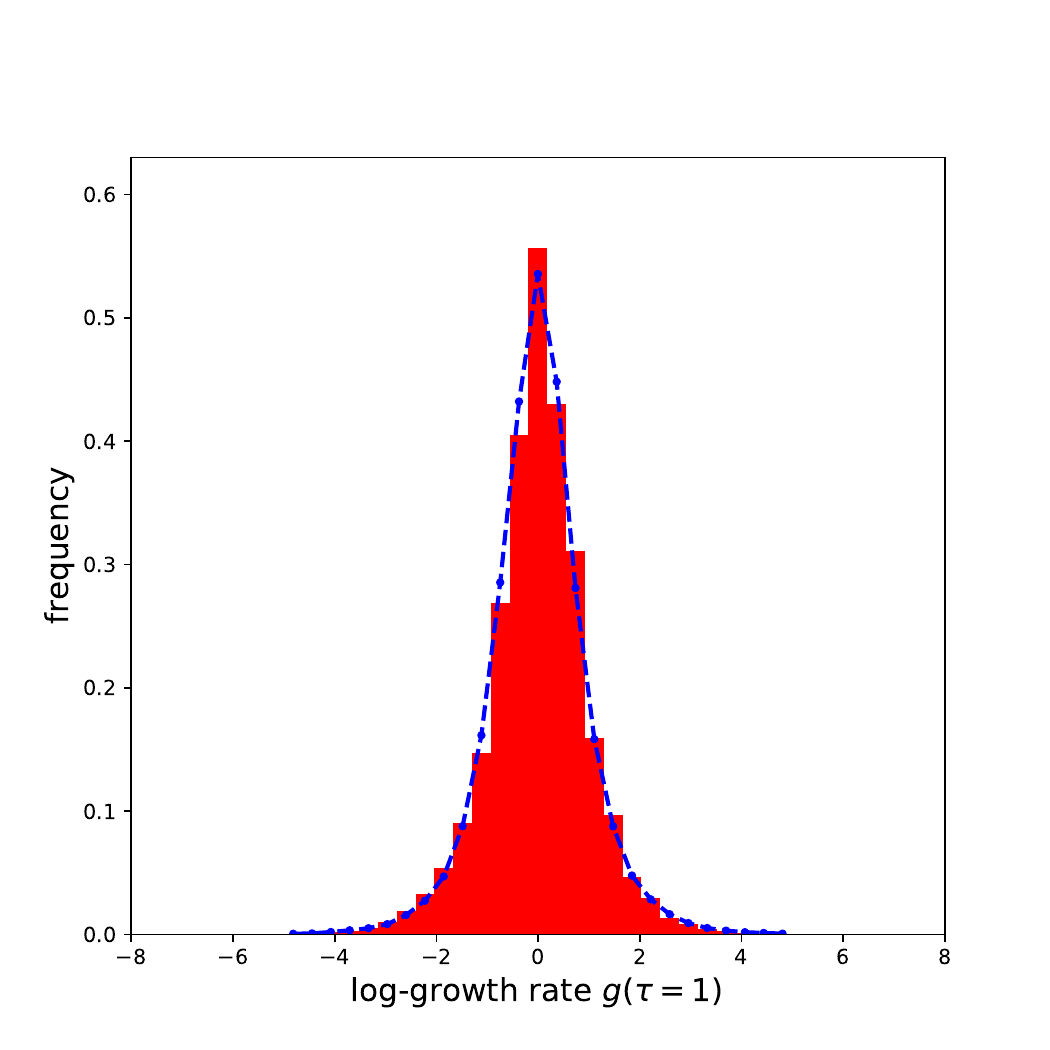}
\includegraphics[angle=0,width=0.4\textwidth]{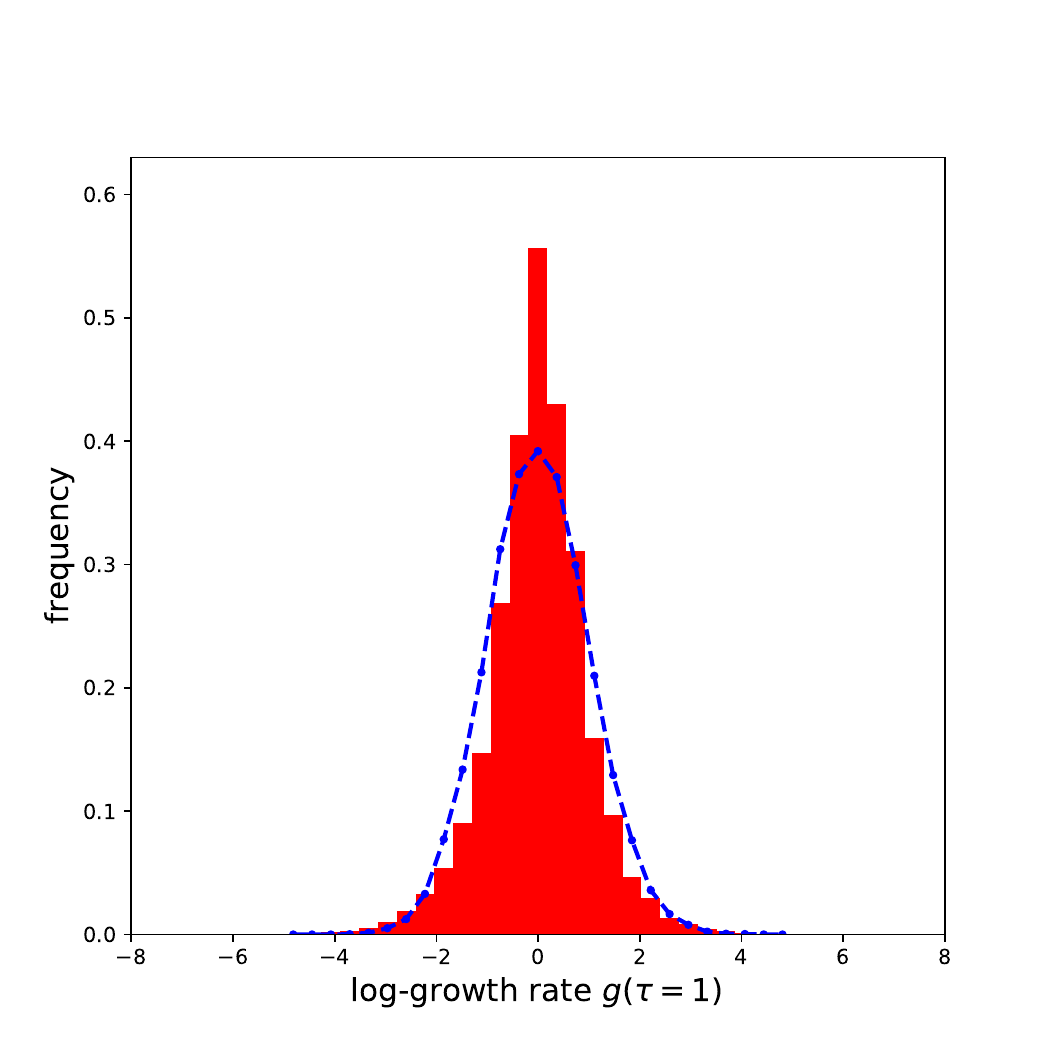}
\end{center}
\caption{\small 
On the left: The red histogram represents the empirical $P(g,\tau=1)$, 
the  blue dashed line shows the distribution
obtained from simulations of the neutral model of eq. \ref{SDE}, 
calibrated by using a scheme based on
the log-growth distribution and the stationary abundance distribution.
On the right: The red histogram represents the empirical $P(g,\tau=1)$, 
the  blue dashed line shows the distribution
obtained from simulations of the logistic model of eq. \ref{eq_Log}. 
In this simulation, $\sigma$ and $K$ parameters were calibrated 
using  the stationary distribution, $a$ using $P(g,\tau=1)$ fitted with equation \ref{eq_DisFin} (see Supplementary Material).
Note that choosing different values of $a$ does not improve the fit with the empirical data.
Fit quality for the two scenarios are compared by measuring 
the Kolmogorov-Smirnov statistic \cite{KS} of the empirical samples 
with the simulated ones.
Simulations using the logistic model yield less accurate results, exhibiting a K-S statistic that is 50\% higher compared to that obtained when employing the SDE described in eq. \ref{SDE}.}
%
\label{Fig_simula}
\end{figure}

Further support to the plausibility of the considered model 
for describing our dataset can be found by 
looking at the $P(g,\tau=1)$ produced 
from long simulated time-series generated by this model at stationarity.
The  SDE of equation \ref{SDE} is calibrated by using 
the $b/D$ and $a$ values obtained from the log-growth distribution
and $D\cdot a$ from the stationary abundance distribution.
In Fig. \ref{Fig_simula} we can see how the simulated 
distribution is comparable to the empirical one.

We can perform the same test using the logistic model of 
equation \ref{eq_Log}. 
As can be seen in  Fig. \ref{Fig_simula}, 
data generated from simulations poorly describe the experimental ones.
Even if the distribution of the simulated data generally match the scale of the dispersion 
of the empirical one,
its shape is not close to the leptokurtic empirical distribution.
This result is in accordance with our 
prediction of the approximation of eq. \ref{eq_LogRet1}, 
which shows that the distribution is always
practically indistinguishable from a Gaussian one.\\

\begin{figure}[h]
\begin{center}
\includegraphics[angle=0,width=0.49\textwidth]{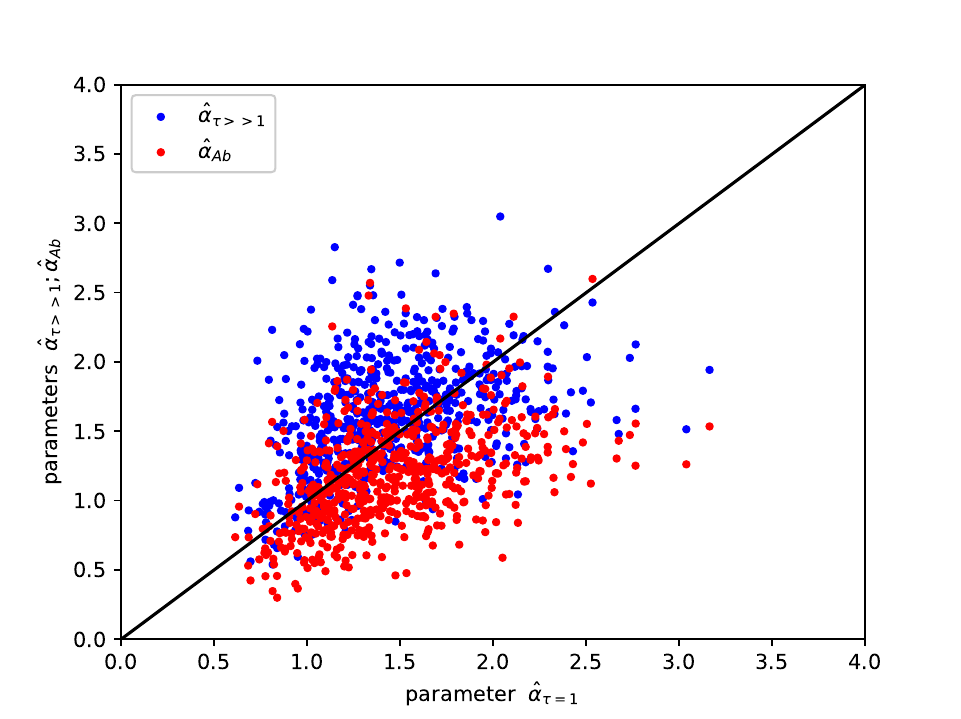}
\includegraphics[angle=0,width=0.49\textwidth]{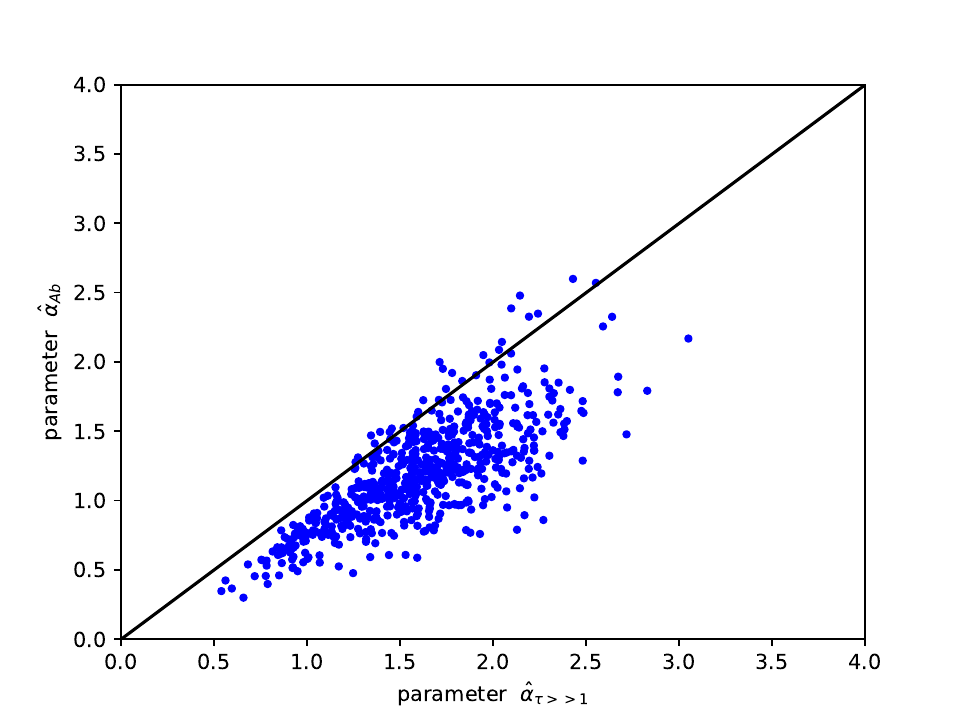}
\end{center}
\caption{\small On the left: In this scatter plot blue points represent
the $\hat{\alpha}_{\tau=1}$ ($\alpha$ estimated from $P(g,\tau=1)$),
versus $\hat{\alpha}_{\tau>>1}$ ($\alpha$ inferred from
$P(g,\tau>>1)$).
Red points stand for $\hat{\alpha}_{\tau=1}$ versus $\hat{\alpha}_{Ab}$
($\alpha$ estimated from the stationary abundance distribution).
The Pearson correlation is equal to 0.40 and 0.43, respectively.
On the right:  The scatter plot of 
$\hat{\alpha}_{\tau>>1}$ versus $\hat{\alpha}_{Ab}$.
The Pearson correlation is equal to 0.76.
The solid lines are $x=y$.
}
\label{Fig_scatter}
\end{figure}



In the following we analyse each time-series taken one by one. 
In this case we do not assume that they are independent, even though the parameters
that we fit are effective, thus possibly including the effect of interactions. 
This analysis allows to verify the plausibility of the neutral hypothesis and to
check the consistency of the different methods used to estimate the parameter 
values using the model of eq. \ref{SDE}. %
In this approach, we independently fit the log-growth rate distribution
for $\tau=1$ (2 fitting parameters: $\alpha$ and $a$), the log-growth rate distribution
for $\tau>>1$ (1 fitting parameter: $\alpha$) and the stationary abundance distribution (2 fitting parameters: $\alpha$ and $D\cdot a$).  
For each time series, we compare the three independently 
estimated values of $\alpha$, which we indicate as 
$\hat{\alpha}_{\tau=1}$, $\hat{\alpha}_{\tau>>1}$ and $\hat{\alpha}_{Ab}$.
Results are shown in Fig. \ref{Fig_scatter}.
Despite the dispersion 
of the points, which accounts for the heterogeneity of the time series, 
the estimated values are 
distributed close to the line $x=y$.
Taking into account the size and noise of the time-series,
the result 
suggests that the different approaches
used to estimate the parameter values are consistent. 
A small systematic deviation between $\hat{\alpha}_{\tau>>1}$ and 
$\hat{\alpha}_{Ab}$
can 
suggests that the estimated 
$\hat{\alpha}_{\tau>>1}$  are in general larger than the $\hat{\alpha}_{Ab}$.

This analysis supports the validity of applying the approximation that
 all OTUs behave according to eq. \ref{SDE} with 
 effective parameters.
In fact, the mean value of $\hat{\alpha}_{\tau=1}$ 
is equal to $1.45\pm0.41$ and the estimation of $a$ coming from the analysis of each time-series taken one by one gives $3.63\pm2.91$ days.
These values are consistent with the ones found by analyzing the ensemble of all time-series together.
Moreover, the coefficients of variation of the distributions of the three $\alpha$ are always smaller or equal to $0.3$, suggesting that the three distributions present a low-variance.
The results presented in Fig. \ref{Fig_scatter} excluded 
the time series which generate a $P(g,\tau=1)$ 
with an excess kurtosis smaller than $0.5$, corresponding 
to the $12\%$ of the original dataset. 
For these data the estimation of $\hat{\alpha}_{\tau=1}$
is inaccurate. 
More details can be found in the Supplementary material.
\\

The analysis on a per-OTU basis was also carried out by using the model defined in eq. \ref{eq_Log}.
Considering all the OTUs, we run the statistical normality test of Shapiro-Wilk 
for each $P(g,\tau=1)$.
This test produces a p-value smaller than 0.05 in $86\%$ of the cases, showing that eq. \ref{eq_Log} can not reproduce these empirical distributions.
The same test applied to OTUs with
an excess kurtosis larger than 0.5 provides a 94\% of fails.

Finally, a model selection approach was used to compare the two models. 
We considered the distribution of eq. 3, generated by the CIR equation, and the Normal one, generated by the logistic SDE. 
To quantify the evidence supporting each model we used the Akaike information criterion (AIC), which compares models likelihoods. The AIC is calculated as follows: $AIC = 2 K- 2 L$, where $K$ is the number of parameters of the model  and $L$ is the maximum log-likelihood. 
The model with the lowest Akaike information is the best supported model \cite{Anderson}. 
The analysis shows that the distribution of eq. 3 is the best supported model for the 
97\% of the considered OTUs. 
Restricting the dataset to OTUs with distributions presenting an excess kurtosis larger than 0.5, equation 3 is always the best supported model. 
Akaike weights are displayed in the Supplementary Material.

\section{Discussion}

Our results show that the considered neutral model with demographic stochasticity
can successfully 
describe the log-growth rate distributions and the stationary abundance distribution
derived from the stationary OTUs abundance time-series.
More significantly, the model can predict the temporal dependence of the 
log-growth rate distribution, by reproducing the kurtosis evolution of $P(g,\tau)$
as a function of $\tau$. 
Furthermore, the typical shape of $P(g,\tau>>1)$ can be independently 
assessed when using this approach. 
We can observe that this last distribution generally has a shape that is relatively close to a Gaussian one. 
It can be hypothesized that 
temporal dependencies 
of $g$ at different times $t$
produce a slight deviation from the convergence to a Gaussian shape, as suggested by aggregational Gaussianity in the case of time additive independent variables.
In contrast, this heuristic reasoning based on the central limit theorem, is not useful for describing the temporal evolution of Var$_P$. If such considerations were valid, at large $\tau$, Var$_P\propto \tau$, and it would not saturate to a limit value, as it is displayed by our analysis. Note that the law Var$_P\propto \tau$ applies to diffusive models used in population dynamics \cite{Holmes}, but if a regulation process is present, for example in the form of a density dependence,
Var$_P \propto \tau^{2H}$ with $H$, the Hurst exponent, smaller than $0.5$. 
This behavior has already been reported since the classical work of Keitt {\it et al.} \cite{Keitt98} and it can be understood by considering that 
regulation processes introduce in $g$ anticorrelations at different times. 
In our case, instead, Var$_P$ grows 
at small $\tau$,  and then saturates to a limit value. 
This is the first time that the idea presented in Kalyuzhny {\it et al.} \cite{Kalyuzhny}, which conjectures
that, for  stabilizing forces which drive the populations towards an equilibrium, Var$_P$ should reach a saturation point, are quantitatively confirmed by empirical data and analytical considerations.

In addition to these results, obtained comparing the analytical predictions of the model with the experimental data, we can produce some numerical results which support these outcomes.
In fact, the SDE, calibrated with the parameters obtained
from the fitting of the distributions, is able to generate simulated processes with a log-growth rate synthetic 
distribution comparable with the empirical one.

Finally, the analysis of each time-series, taken one by one, confirms the consistency of the parameters inferred by independently fitting the log-growth rate distribution, the same distribution for large $\tau$ and the stationary abundance distribution.
The coefficients of variation of the distributions of these parameters are relatively small and the comparison of their mean values with the estimations derived from the analysis of the ensemble of all time-series together are also consistent. 
These facts support an effective neutral modeling approach
in a first approximation.\\

The most relevant result of our work is the description of the subtle temporal dependence of the log-growth rate distribution.
The importance of this result is due to the fact that the distributions that reproduce the abundance and growth rate are flexible enough for describing very different datasets.
On varying its parameters, distinct distributions can be seen as 
a special case of the Gamma one, which can describe data which present shapes close to
power-laws, exponential, and even log-normals.
The expression of eq. \ref{ana_growth} 
turned out to be really versatile 
and has been able to reproduce 
log-growth rate generated by very different systems, as can be seen in \cite{Ashish}. 
The 
 fit of these distributions 
is important, but not necessarily conclusive
for claiming that the considered SDE, with a mean reverting linear drift and demographic noise, can account for the description of so different datasets \cite{Ashish}. 
Other features or more specific characteristics of the considered distribution should be assessed. This is achieved in our study by analyzing the temporal dependence
of the $P(g,\tau)$.\\ 


Another important point raised by our analysis is the fact that 
the logistic model with an environmental stochastic term is not suited for 
describing the $P(g,\tau)$ found in the considered microbiota dataset at stationarity.
Our analytical results demonstrated that for the regimes with small $\sigma$ 
this process produces normal $P(g,\tau)$ with a variance dependent on $\sigma$, $a$ and $\tau$. Numerical simulations confirmed that this analytical approximation is 
good in all the regimes. 
For these reasons, this model 
can not reproduce empirical log-growth rates with leptokurtic distributions, as for the analysed data,  which present a large and positive kurtosis. 
This is confirmed by statistical normality tests, which rejected the normality hypothesis.
The same test was performed for each individual OTUs, 
excluding the normality hypothesis in the vast majority of cases.
Finally, a model selection approach showed that the Normal distribution is the best supported model, in relation to the model of equation 3, for only 3\% of the OTUs.
Note that these results only show that, at least in the current formulation, there is no support for using the $P(g,\tau)$ produced by the logistic model with a linear stochastic term. These results should not 
be necessarily interpreted as evidence for eq. 1 to be more appropriate 
than eq. 5 for describing the microbiome population dynamics.
This is not  the aim of this work and, in order to attempt to answer this question, further analyses, richer and more robust empirical data are needed.  
\\

Finally, our study suggests
that neutral models can effectively describe the population dynamics
of bacteria in the considered microbiota. 
The existent literature on this theme reported conflicting evidence.
It suggests that human microbial communities are not generally neutral 
but a small minority of cases already demonstrated the existence of neutral processes \cite{Li16,eLife}.
These assessments were generally obtained carrying out an analysis of the characteristic of the community ecology  based on macroecological statistical properties. 
Here, we arrived at this conclusions using a dynamical population approach.
In this sense, our results, even if obtained over a limited dataset, are  
particularly interesting for their implications at the level of biological factors that
control the dynamics of the considered populations.
The models we have analysed in this paper seem to suggest that
demographic noise 
is relatively more important than the environmental one for explaining $P(g,\tau)$.
Stochastic logistic models with quenched noise (e.g., random parameters) or which 
encompass time-correlated stochastic terms can provide a wider variability in the temporal evolution of the population size and therefore an improved explanatory power which might better describe log-growth distributions. However, these approaches are outside the scope of the present work.




\section*{Data Availability}

This paper does not use original data. The datasets analysed are available from the original references which are listed in the manuscript. E.B. can be contacted to request the processed data from this study or they can be downloaded at:  
https://figshare.com/s/cf765721ff4f6ff7d92a.

\section*{Acknowledgments}

E.B. thanks the LIPh lab at the Physics Department of the 
Padova University for its hospitality during the realization of this work.
The authors acknowledge Jacopo Pasqualini for pre-processing the dataset considered in this analysis and Amos Maritan, Emanuele Pigani and Samir Suweis for fruitful discussions.

E.B. received partial financial support from the National Council for Scientific and Technological Development - CNPq (Grant No. 305008/2021-8) and FAPERJ (Grant No. 260003/005762/2024).
S.A. acknowledges financial support under the National Recovery and Resilience Plan (NRRP), Mission 4, Component 2, Investment 1.1, Call for tender No. 104 published on 2.2.2022 by the Italian Ministry of University and Research (MUR), funded by the European Union - NextGenerationEU - Project Title Emergent Dynamical Patterns of Disordered Systems with Applications to Natural Communities - CUP 2022WPHMXK - Grant Assignment Decree No. 2022WPHMXK adopted on 19/09/2023 by the Italian Ministry of Ministry of University and Research (MUR).


\end{document}